\documentclass[reprint,superscriptaddress,aps,twocolumn,showpacs]{revtex4-1}
\usepackage{amssymb}
\usepackage{amsmath}
\usepackage{graphicx}
\usepackage{multirow}
\usepackage{hyperref}
\usepackage{color}
\usepackage{gensymb}

\newcommand{\splitcell}[1]{\begin{tabular}{@{}c@{}}#1\end{tabular}}
\newcommand{\bsplitcell}[1]{$\bigg(\splitcell{#1}\bigg)$}

\allowdisplaybreaks

\begin{document}
\title{Mass spectrum of $1^{--}$ heavy quarkonium}

\author{Zheng Zhao}
\email[]{zhaozheng1022@hotmail.com}
\author{Kai Xu}
\email[]{gxukai1123@gmail.com}
\affiliation{School of Physics and Center of Excellence in High Energy Physics and Astrophysics, Suranaree University of Technology, Nakhon Ratchasima 30000, Thailand}
\affiliation{China-Thailand Joint Research Center of Physics, Harbin Engineering University, People's Republic of China and Suranaree University of Technology, Nakhon Ratchasima 30000, Thailand}
\author{Ayut Limphirat}
\email[]{ayut@g.sut.ac.th}
\affiliation{School of Physics and Center of Excellence in High Energy Physics and Astrophysics, Suranaree University of Technology, Nakhon Ratchasima 30000, Thailand}
\affiliation{China-Thailand Joint Research Center of Physics, Harbin Engineering University, People's Republic of China and Suranaree University of Technology, Nakhon Ratchasima 30000, Thailand}
\author{Warintorn Sreethawong}
\email[]{warintorn.sut@gmail.com}
\affiliation{School of Physics and Center of Excellence in High Energy Physics and Astrophysics, Suranaree University of Technology, Nakhon Ratchasima 30000, Thailand}
\author{Nattapat Tagsinsit}
\author{Attaphon Kaewsnod}
\affiliation{School of Physics and Center of Excellence in High Energy Physics and Astrophysics, Suranaree University of Technology, Nakhon Ratchasima 30000, Thailand}
\author{Xuyang Liu}
\affiliation{School of Physics and Center of Excellence in High Energy Physics and Astrophysics, Suranaree University of Technology, Nakhon Ratchasima 30000, Thailand}
\affiliation{School of Physics, Liaoning University, Shenyang 110036, China}
\author{Khanchai Khosonthongkee}
\affiliation{School of Physics and Center of Excellence in High Energy Physics and Astrophysics, Suranaree University of Technology, Nakhon Ratchasima 30000, Thailand}
\author{Sampart Cheedket}
\affiliation{Department of Physics, School of Science, Walailak University, Nakhon Si Thammarat,  80160, Thailand}
\author{Yupeng Yan}
\email[]{yupeng@g.sut.ac.th}
\affiliation{School of Physics and Center of Excellence in High Energy Physics and Astrophysics, Suranaree University of Technology, Nakhon Ratchasima 30000, Thailand}
\affiliation{China-Thailand Joint Research Center of Physics, Harbin Engineering University, People's Republic of China and Suranaree University of Technology, Nakhon Ratchasima 30000, Thailand}

\date{\today}

\begin{abstract}
	\indent We calculate the masses and leptonic decay widths of the bottomonium $b\bar b$ and charmonium $c\bar c$ states in a constituent quark model where the Cornell-like potential and spin-dependent interaction are employed, with all model parameters predetermined by studying ground and first radial excited states of S- and P-wave heavy quarkonium mesons.
	By comparing the theoretical predictions for $J^{PC}=1^{--}$ quarkonium states with experimental data and considering possible mixtures of $nS$ and $(n-1)D$ states, we provide tentative assignments for all observed $J^{PC}=1^{--}$ heavy quarkonia. The work suggests that the $\Upsilon$(10860) and $\Upsilon$(11020) are $b\bar b$ $5S-4D$ mixture states, and the $\psi$(4360) and $\psi$(4415) are largely $4S$ and $3D$ $c\bar c$ states respectively. The $\psi$(4230) may not be accommodated with the conventional meson picture in the present work.
\end{abstract}

\maketitle

\section{Introduction}\label{sec:Int}

Over 20 charmoniumlike and bottomoniumlike XYZ states have been observed in the past two decades. The charged states (Z states), which might be good exotic state candidates in the tetraquark or molecule picture, have inspired extensive interests of theorists in revealing their underlying structures~\cite{Chen:2016qju}. However, distinguishing the real exotic neutral X and Y states from conventional meson states is still a challenging work, and underlying structures of X and Y states are still wildly discussed and debated in the past decade~\cite{Eichten:2007qx,Liu:2019zoy}.

The exotic states with $J^{PC}=1^{--}$, also known as Y states, are named $\Upsilon$ in the bottomonium region, and $\psi$ in the charmonium region according to the latest PDG (Particle Data Group) naming scheme~\cite{PDG}. It is significant to separate these exotic neutral states from conventional meson picture before treating them in other exotic pictures.

In the bottomonium region, the PDG states $\Upsilon$(9460), $\Upsilon$(10023), $\Upsilon$(10355), and $\Upsilon$(10579) are assigned to be $\Upsilon$(1S) to $\Upsilon$(4S) respectively~\cite{PDG}. Meanwhile, in the charmonium region, the $J/\psi$, $\psi$(3686), $\psi$(3770), $\psi$(4040), $\psi$(4160) are assigned to be $\psi(1S)$, $\psi(2S)$~\cite{PDG}, $\psi(1D)$~\cite{Zhang:2006td,Li:2009zu,Segovia:2008zz,Barnes:2005pb}, $\psi(3S)$~\cite{Zhang:2006td,Li:2009zu,Segovia:2008zz,Barnes:2005pb}, and $\psi(2D)$~\cite{Zhang:2006td,Li:2009zu,Segovia:2008zz,Barnes:2005pb} respectively. 
Theoretical pictures of $c\bar c$ bound states including S-D mixings~\cite{Li:2009zu,Shah:2012js,Wang:2019mhs,Wang:2022jxj}, hybrid charmonium $c\bar cg$~\cite{HadronSpectrum:2012gic,Luo:2005zg,Zhu:2005hp}, compact tetraquark $\left(qc\bar q\bar c\right)$~\cite{Maiani:2005pe,Drenska:2009cd,Ebert:2008kb}, and molecule $\left(q\bar c\right)\left(\bar qc\right)$~\cite{Chiu:2005ey,Ding:2008gr,Close:2009ag,Close:2010wq} have been proposed for studying the higher excited states, $\psi$(4230), $\psi$(4360), $\psi$(4660), and $Y(4500)$ observed by BESIII recently~\cite{BESIII:2022joj}.

Meanwhile, experimental new values of mass and leptonic decay width have been reported for these $J^{PC}=1^{--}$ states, and the understanding of these states has been also improved since many theoretical works have been done. However, theoretical predictions of leptonic widths for higher excited states are still not consistent with the latest experimental data~\cite{Shah:2012js,Godfrey:2015dia,Wang:2018rjg,Segovia:2016xqb,Li:2009zu,Segovia:2008zz,Barnes:2005pb,Barnes:2004cz,Mutuk:2018xay}.
All established heavy quarkonium states with $J^{PC}=1^{--}$ are listed in Table~\ref{tab:1--exp}, with experimental data of mass and leptonic decay width from PDG~\cite{PDG}, and also with assignments from cited theoretical works. 
We briefly review the model of those works here and discuss their results in Sec.~\ref{sec:DIS}.

Masses and leptonic widths of heavy quarkonium are estimated in a Martin-like potential model where a non-Coulombic power law potential is employed~\cite{Shah:2012js}. Different parameters are applied for bottomonium and charmonium mesons. 

In Refs~\cite{Godfrey:2015dia,Wang:2018rjg}, masses and decay properties of excited bottomonium states are studied in a relativized quark model (Godfrey-Isgur model) developed from Refs.~\cite{Godfrey:1985xj,Godfrey:1985by,Godfrey:1985sp,Godfrey:2004ya,Godfrey:2005ww} where a Cornell-like potential is employed. Masses, radiative transitions, annihilation decays, hadronic transitions and production cross sections of excited bottomonium states are evaluated. 

Bottomonium mass spectrum, electromagnetic, strong and hadronic decays are also studied in a non-relativistic quark model~\cite{Segovia:2016xqb} developed from~\cite{Vijande:2004he} and their previous work~\cite{Segovia:2008zza}.

Charmonium spectrum, and electromagnetic decays are estimated in a non-relativistic model with a coulomb potential plus a screened linear potential~\cite{Li:2009zu}, and also are studied in a constituent quark model with a screened confinement potential~\cite{Segovia:2008zz}. In Ref~\cite{Barnes:2005pb}, higher charmonium mass spectra are calculated in a non-relativistic model with a Cornell-like potential, and the corresponding leptonic widths are estimated in Ref~\cite{Barnes:2004cz}. 

\begin{table}[tb]
\caption{Mass and leptonic decay width of bottomoium and charmonium $1^{--}$ states from PDG~\cite{PDG}, and their assignments from cited sources.}
\label{tab:1--exp}
\begin{ruledtabular}
\begin{tabular}{lcccccccc}
State & $M^{exp}  {\rm (MeV)}$ & $\Gamma^{exp} {\rm (keV)}$& Assignment
\\
\hline
 $\Upsilon$(1S)& $9460$ & $1.340\pm0.018$                               & 1S $b\bar b$~\cite{PDG}
\\
$\Upsilon$(2S)& $10023$ & $0.612\pm0.011$                              & 2S $b\bar b$~\cite{PDG} 
\\
$\Upsilon$(3S)& $10355$ & $0.443\pm0.008$                              & 3S $b\bar b$~\cite{PDG}
\\
$\Upsilon$(4S)& $10579$ & $0.272\pm0.029$                              & 4S $b\bar b$~\cite{PDG}
\\
$\Upsilon$(10860)& $10885.2^{+2.6}_{-1.6}$ & $0.31\pm0.07$   & 5S $b\bar b$~\cite{Godfrey:2015dia,Wang:2018rjg,Segovia:2016xqb,Deng:2016ktl,Dong:2020tdw}
\\
$\Upsilon$(11020)& $11000\pm4$ & $0.13\pm0.03$                     & 6S~\cite{Godfrey:2015dia,Wang:2018rjg,Segovia:2016xqb,Deng:2016ktl,Dong:2020tdw} 
\\
&&& 7S $b\bar b$~\cite{Shah:2012js}
\\
 $\psi$(1S)& $3097$ & $5.55\pm0.14$& 1S $c\bar c$~\cite{PDG}
\\
$\psi$(2S)& $3686$ & $2.33\pm0.04$ & 2S $c\bar c$~\cite{PDG}
\\
$\psi$(3770)& $3773$ & $0.26\pm0.02$ & 1D $c\bar c$~\cite{HadronSpectrum:2012gic,Zhang:2006td,Li:2009zu,Segovia:2008zz,Barnes:2005pb}
\\
$\psi$(4040)& $4039\pm1$ & $0.86\pm0.07$ & 3S $c\bar c$~\cite{Zhang:2006td,Segovia:2008zz,Li:2009zu,Barnes:2005pb}
\\
$\psi$(4160)& $4191\pm5$ & $0.48\pm0.22$ & 2D $c\bar c$~\cite{Zhang:2006td,Segovia:2008zz,Li:2009zu,Barnes:2005pb}
\\
$\psi$(4230)& $4230\pm8$ & ... & 4S $c\bar c$~\cite{Llanes-Estrada:2005qvr, Li:2009zu, Shah:2012js}
\\
&&&3D $c\bar c$~\cite{Zhang:2006td,Dai:2012pb}
\\
&&&$c\bar cg$~\cite{HadronSpectrum:2012gic,Luo:2005zg,Zhu:2005hp}
\\
&&&$\left(qc\bar q\bar c\right)$~\cite{Maiani:2005pe,Drenska:2009cd,Ebert:2008kb}
\\
&&&$\left(q\bar c\right)\left(\bar qc\right)$~\cite{Chiu:2005ey,Ding:2008gr,Close:2009ag,Close:2010wq}
\\
$\psi$(4360)& $4368\pm13$ & ...                   & 4S $c\bar c$~\cite{Segovia:2008zz}, 3D $c\bar c$~\cite{Li:2009zu}
\\
$\psi$(4415)& $4421\pm4$ & $0.58\pm0.07$ & 4S $c\bar c$~\cite{Zhang:2006td}, 3D $c\bar c$~\cite{Segovia:2008zz}
\\
                    &                      &                          & 5S $c\bar c$~\cite{Li:2009zu,Shah:2012js}
\\
$\psi$(4660)& $4643\pm9$ & ...                      & 5S $c\bar c$~\cite{Segovia:2008zz,Ding:2007rg}
\\
                    &                      &                          & 6S $c\bar c$~\cite{Li:2009zu,Shah:2012js}
\end{tabular}
\end{ruledtabular}
\end{table}

In this work, we apply a model developed from Ref.~\cite{Zhao:2021bma, Zhao:2021jss} to predict the masses and leptonic decay widths of higher excited $1^{--}$ bottomonium $b\bar b$ and charmonium $c\bar c$ states. By considering possible S-D mixtures, and comparing the theoretical results with experimental data, we present possible conventional meson interpretation for the higher excited $1^{--}$ heavy quarkonium states. The states which can not be accommodated in the present picture will be studied in our future work by applying exotic pictures.

The paper is organized as follows. In Sec.~\ref{sec:TM}, a constituent quark model~\cite{Zhao:2021bma, Zhao:2021cdv, Zhao:2021jss,Xu:2019fjt, Xu:2020ppr} is developed to include a spin-dependent interaction~\cite{Schoberl:1986bv} for studying higher orbital excited quarkonium states. In Sec.~\ref{sec:DIS}, theoretical masses and leptonic decay widths of the $1^{--}$ heavy quarkonium states are calculated and compared with experimental data. Tentative assignments for higher excited heavy quarkonium states are suggested in S-D mixture picture. A summary is given in Sec.~\ref{sec:SUM}.

\section{\label{sec:TM}THEORETICAL MODEL}

\indent The non-relativistic Hamiltonian for studying the meson system takes the form,
\begin{flalign}\label{eqn:ham}
H &= H_0+ H_{SD}, 
\end{flalign}
with
\begin{flalign}\label{eqn:h0}
H_{0}& = M_{ave}+\frac{p^2}{2m_{r}}+(A r-\frac{B}{r}), \nonumber \\
H_{SD}&= C_{SS}(r)\vec \sigma_1\cdot\vec \sigma_2+C_{LS}(r)\vec L\cdot\vec S +C_{T}(r)S_{12},
\end{flalign}
where {$H_0$} is taken from the previous work~\cite{Zhao:2021bma, Zhao:2021jss, Zhao:2021cdv}. $\vec r$ is the relative coordinate between the two quarks, $M_{ave}$ is the spin-averaged mass taken from experimental data~\cite{PDG}, and $m_{r}$ stands for the reduced quark mass taking the form $m_1m_2/(m_1+m_2)$. In the work, we employ $m_c = 1270 \ {\rm MeV}$ and $m_b = 4180 \ {\rm MeV}$~\cite{PDG}. $\vec L$, $\vec S$, and $\vec J$ are the operators of orbital angular momentum, total spin, and total angular momentum, respectively. The tensor operator $S_{12}$ is defined as
$S_{12}=(3(\vec \sigma_1\cdot r)(\vec \sigma_2\cdot r)-\vec \sigma_1\cdot\vec \sigma_2)$. 

$C_{SS}(r)$, $C_{LS}(r)$, and $C_{T}(r)$ in  Eq. (\ref{eqn:h0}) are derived by following the Breit-Fermi interaction, that is,  
\begin{flalign}\label{eqn:BF}
C_{SS}(r)&=\frac{1}{6m_i^2}\Delta V_V(r)=\frac{2B{\sigma}^3e^{-{\sigma}^2r^2}}{3\sqrt \pi m_i^2}, \nonumber \\
C_{LS}(r)&=\frac{1}{2m_i^2}\frac{1}{r}[3\frac{dV_V(r)}{dr}-\frac{dV_S(r)}{dr}] \nonumber \\
&=-\frac{A}{2m_i^2}\frac1r-\frac{3B\sigma}{\sqrt \pi m_i^2}\frac{e^{-{\sigma}^2r^2}}{r^2}+\frac{3B}{2m_i^2}\frac{Erf[\sigma r]}{r^3},
 \nonumber \\
C_{T}(r)&=\frac1{12m_i^2}[\frac1r\frac{dV_V(r)}{dr}-\frac{d^2V_V(r)}{dr^2}] \nonumber \\
&=-\frac{B\sigma e^{-{\sigma}^2r^2}}{2\sqrt \pi m_i^2 r^2}-\frac{B{\sigma}^3e^{-{\sigma}^2r^2}}{3\sqrt \pi m_i^2}+\frac{B Erf[\sigma r]}{4m_i^2 r^3}, 
\end{flalign}
Note that we have employed $V_V(r)=-B\,Erf[\sigma r]/r$ and $V_S(r)=Ar$, taken from Ref.~\cite{Schoberl:1986bv}.

$\vec\sigma_i$ in Eq. (\ref{eqn:h0}) are quark spin operators, and the contribution of $\vec\sigma_{i}\cdot\vec\sigma_{j}$ is $-3$ for $S=0$ and $+1$ for $S=1$ mesons. The matrix elements of $\vec L\cdot\vec S$ and $S_{12}$ in the $|JMLS\rangle$ basis read
\begin{flalign}
&\langle\vec L\cdot\vec S\rangle=[J(J+1)-L(L+1)-S(S+1)]/2, \nonumber \\
&\langle S_{12}\rangle = \left\{ \begin{array}{ll}
-\frac {2L}{2L+3} & \textrm{$J=L+1$}\\
+2 & \textrm{$J=L$}\\
-\frac{2(L+1)}{2L-1} & \textrm{$J=L-1$}
\end{array} \right.
\end{flalign}
The tensor operator $S_{12}$ has non-vanishing matrix elements between the two orbital parts of spin-triplet states. 

\indent The string tension coefficient $A$ and Coulomb coefficient $B$ in Cornell potential $V(r)=Ar-B/r $ may take different values when $A$ and $B$ are fitted to charmonium and bottomonium experimental data. This indicates that $A$ and $B$ might be flavor dependent parameters. Inspired by lattice QCD studies~\cite{Kawanai:2011xb, Ikeda:2011bs}, $A$ and $B$ are proposed to be mass dependent coupling parameters. For more detailed discussion, one may refer to Ref.~\cite{Zhao:2021bma}. The hyperfine coefficient $\sigma$ is also proposed to be mass dependent~\cite{Schoberl:1986bv}. 

\indent In this work, parameters $A$, $B$ and $\sigma$ are assumed to take the following mass dependent form
\begin{eqnarray}
A= a+bm_{i},\;B=B_0 \sqrt{\frac{1}{m_{i}}},\;\sigma =\sigma_0 m_{i},
\end{eqnarray}
with $a$, $b$, $B_0$, and $\sigma_{0}$ being constants. Four model coupling constants are determined by comparing the theoretical mass results with experimental data of conventional mesons,
\begin{eqnarray}\label{eq:nmo1}
&a=78650 \ {\rm MeV^2}, \quad b=28 \ {\rm MeV}\,,  \nonumber\\
&B_0=30.86 \ {\rm MeV^{1/2}}\,, \quad \sigma_0=0.7\,.
\end{eqnarray}

The fitting results $M^{cal}$ for the S- and P-wave ground and first radial excited bottomonium and charmonium meson states are listed in Table~\ref{tab:mass}, together with experimental data $M^{exp}$ from PDG~\cite{PDG}. Some typical theoretical mass results from other works for bottomonium mesons~\cite{Segovia:2016xqb,Godfrey:2015dia,Shah:2012js} and charmonium mesons~\cite{Barnes:2005pb,Li:2009zu} are collected for comparison. Our results are fairly compatible with experimental data.

\begin{table}[tb]
\caption{Masses of ground and first radial excited bottomonium and charmonium meson states, with unit in MeV. $M^{cal}$ are our fitting results. Experimental data $M^{exp}$ are taken from PDG~\cite{PDG}, and other theoretical results for comparison are from cited sources.}
\label{tab:mass}
\begin{ruledtabular}
\begin{tabular}{lcccccccc}
$b\bar b$ & $J^{PC}$ & $nL$ & $M^{exp}$ & $M^{cal}$ & \cite{Segovia:2016xqb} & \cite{Godfrey:2015dia} & \cite{Shah:2012js}
\\
\hline
   $\eta_b$ & $0^{-+}$ & 1S &  9399  &  9394 & 9455   & 9402 & 9392
 \\
                  &               & 2S  &  9999 &  9989 & 9990   & 9976 & 9991
 \\
$\Upsilon$ & $1^{--}$ & 1S  &  9460 &  9461 & 9502   & 9465 & 9460
\\
                  &              & 2S & 10023 & 10017 & 10015 & 10003&10024
\\
       $h_b$ &$1^{+-}$ & 1P & 9899   & 9894   & 9879   & 9882 & 9896
\\
                  &              & 2P & 10260 & 10270 & 10240 & 10250&10260
\\
$\chi_{b0}$&$0^{++}$& 1P & 9859   &  9859  & 9855  & 9847  & 9862
\\
                  &              & 2P & 10232 & 10244 & 10221 & 10226&10240
\\
$\chi_{b1}$&$1^{++}$ &1P & 9893   &  9888  & 9874  & 9876  &9888
\\
                  &              & 2P & 10255 & 10266 & 10236 &10261 &10256
\\
$\chi_{b2}$&$2^{++}$&1P &    9912 & 9905   & 9886  & 9897 & 9908
\\
                  &              & 2P&  10269 & 10280 & 10246 & 10261&10268
\\
\hline
$c\bar c$ & $J^{PC}$ & $nL$ & $M^{exp}$ & $M^{cal}$ & \cite{Barnes:2005pb}NR & \cite{Barnes:2005pb}GI & \cite{Li:2009zu}
\\
\hline
 $\eta_c$   & $0^{-+}$&1S &  2984  & 2987 & 2982 & 2975 & 2979
 \\
                  &              &2S &   3638 & 3633 & 3630 & 3623 & 3623
 \\
 $\psi$       & $1^{--}$& 1S &   3097 & 3110 & 3090 & 3098 & 3097
\\
                 &               &2S &   3686 & 3673 & 3672 & 3676 & 3673
\\
 $h_c$      & $1^{+-}$&1P &    3525 & 3533 & 3516 & 3517 & 3519
\\
 $\chi_{c0}$& $0^{++}$&1P& 3415  & 3460 & 3424 & 3445 & 3433
\\
                 &              & 2P & 3860   & 3884 & 3852 & 3916 & 3842
\\
 $\chi_{c1}$& $1^{++}$ &1P& 3510& 3528 & 3505 & 3510 & 3510
\\
 $\chi_{c2}$ & $2^{++}$ & 1P & 3556 & 3566 &3556&3550& 3554
\\
                  &                & 2P & 3930 & 3949 &3972 &3979 & 3537
\\

\end{tabular}
\end{ruledtabular}
\end{table}

The S-wave and D-wave $1^{--}$ quarkonium leptonic decay width given by the Van Royen-Weisskopf formula~\cite{VanRoyen:1967nq}, including radiative QCD corrections for S-wave~\cite{Barbieri:1979be}, takes the same form with~\cite{Segovia:2016xqb} 
\begin{flalign}\label{eqn:DW}
\Gamma(n^3S_1\to e^+e^- )&=\frac{4\alpha^2e_q^2|R_{nS}(0)|^2}{M_n^2}(1-\frac{16\alpha_s}{3\pi}),  \nonumber\\
\Gamma(n^3D_1\to e^+e^- )&=\frac{25\alpha^2e_q^2|R^{''}_{nD}(0)|^2}{2m_i^4M_n^2}, 
\end{flalign}
where the fine-structure constant $\alpha\simeq 1/137$. $e_q$ is the charge of quarks, $M_n$ is the mass of the decaying quarkonium states, $R_{nS}(0)$ and $R_{nD}(0)$ are the radial wave functions of the $^3S_1$ and $^3D_1$ states at the origin respectively. $\alpha_s$ is the running strong coupling constant, where $\alpha_s(b\bar b)=0.118$ for bottomonium~\cite{Shah:2012js} and $\alpha_s(c\bar c)=0.26$ for charmonium~\cite{Li:2009zu}.

{The difference between performing full integration for leptonic width and applying the lowest order approximation is about 50\% for light mesons, but is about 10\% for charmonium mesons and 4\% for bottomonium mesons. Thus the Van's formula with the first order approximation is reliable to be employed for estimating heavy quarkonium leptonic widths.}

\section{\label{sec:DIS}Results and discussion}

\subsection{\label{sec:mlm} Masses and leptonic widths}

\begin{table*}[htbp]
\caption{Present predictions of bottomoium $b\bar b$ and charmonium $c\bar c$ $1^{--}$ state masses (MeV) and leptonic widths (keV) compared with experimental data from PDG~\cite{PDG} and others theoretical works from cited sources.}
\label{tab:1--md}
\begin{ruledtabular}
\begin{tabular}{cccccccccccccc}
 $nL$($b\bar b$) &$M^{exp}$(MeV) & $M^{cal}$(MeV) &\cite{Shah:2012js} &\cite{Godfrey:2015dia} &\cite{Wang:2018rjg}& \cite{Segovia:2016xqb} & $\Gamma^{exp}$(keV) & $\Gamma^{cal}$(keV)  &\cite{Shah:2012js} &\cite{Godfrey:2015dia} &\cite{Wang:2018rjg}& \cite{Segovia:2016xqb}
\\
\hline
1S &  $9460.30\pm0.26$  &   9461 &  9460  &   9465 & 9463   &   9502  &  $1.340\pm0.018$ & 1.370 & 1.203 & 1.44 & 1.650 & 0.71
\\
2S & $10023.26\pm0.31$ & 10017 & 10024 & 10003 & 10017 & 10015  &  $0.612\pm0.011$ & 0.626 & 0.519 & 0.73 & 0.821 & 0.37
\\
1D & ...                              & 10143 & 10147 & 10138 & 10153 & 10117  & ...                           & 0.002 &...        & 0.001& 0.002& 0.001
\\
3S & $10355.2\pm0.5$    & 10379  & 10346 & 10354 & 10356  & 10349  & $0.443\pm0.008$ & 0.468 & 0.330 & 0.53 & 0.569 & 0.27
\\
2D & ...                             & 10461 & 10427 & 10441 & 10442  & 10414  & ...                           & 0.003 &...        & 0.002& 0.003& 0.003
\\
4S & $10579.4\pm1.2$    & 10678 & 10576 & 10635 & 10612  & 10607  & $0.272\pm0.029$ & 0.393 & 0.242 & 0.39 & 0.431 & 0.21
\\
3D & ...                             & 10739 & 10637 & 10698  & 10675 & 10653  & ...                          & 0.005 &...        & 0.002&0.003&...        
\\
5S & ...                             & 10942 & 10755 & 10878 & 10822  & 10818  & ...                          & 0.346 & 0.191 & 0.33 & 0.348 & 0.18
\\
4D & ...                             & 10991 & 10805 & 10928 & 10871  & 10853  & ...                          & 0.006  &...        & 0.002&0.003&...        
\\
6S & ...                             & 11184 & 10904 & 11102 & 11001  & 10995   & ...                          & 0.313 & 0.158 & 0.27 & 0.286 & 0.15
\\
5D & ...                            & 11224 & 10946 & ...       & 11041   & 11023   & ...                           & 0.008 &...        & ...    &0.003&...        
\\
\hline
 $nL$($c\bar c$)  &$M^{exp}$(MeV) & $M^{cal}$(MeV) &\cite{Shah:2012js} & \cite{Li:2009zu} & \cite{Segovia:2008zz} & \cite{Barnes:2005pb} & $\Gamma^{exp}$(keV) & $\Gamma^{cal}$(keV) &\cite{Shah:2012js} & \cite{Li:2009zu} & \cite{Segovia:2008zz} & \cite{Barnes:2004cz}
\\
\hline
1S & $3096.90\pm0.01$ &   3110 &  3097   &   3097 &   3096 &   3090 & $5.55\pm0.14$ & 6.02  & 4.95 & 6.60 & 3.93 & 12.13
\\
2S & $3686.10\pm0.03$ &   3673 &  3690   &   3673&   3703 &   3672  & $2.33\pm0.04$& 2.33   & 1.69 & 2.40 & 1.78 & 5.03
\\
1D & $3773.13\pm0.35$ &   3782 &  3729   &   3787 &   3796 &   3785 & $0.26\pm0.02$ & 0.14  & ...     & 0.03 & 0.22 & 0.06
\\
3S &     $4039\pm1$       &   4046 &  4030   &   4022 &   4097 &   4072 & $0.86\pm0.07$ & 1.55  & 0.96 & 1.42 & 1.11 & 3.48
\\
2D & ...                            &   4114 &  4056   &   4089 &   4153 &   4142 & ...                      &  0.22 & ...     & 0.04 & 0.30 & 0.10
\\
4S & ...                            &   4355 &  4273  &   4273  &   4389 &   4406  & ...                      & 1.19  & 0.65 & 0.97& 0.78 & 2.63
\\
3D & ...                           &    4404 &  4293   &  4317  &  4426  &...         & ...                      &  0.26 & ...     & 0.04 & 0.33 & ...
\\
5S & ...                           &   4628  &  4464   &  4463  &  4614  & ...        & ...                      &  0.97 & 0.49 & 0.70 & 0.57 & ...
\\
4D & ...                           &   4667  & 4480   &...          &  4641  &...         & ...                      & 0.20 & ...      &...      & 0.31 & ...
\\
6S & ...                           &   4879  & 4622   & 4608    &  4791  & ...        & ...                      & 0.82 & 0.39  & 0.49 & 0.42 & ...
\\
5D & ...                           &   4910  & 4634   &...          &  4810  &...         & ...                      & 0.23 & ...      &...      & 0.28 & ...
\\
\end{tabular}
\end{ruledtabular}
\end{table*}

We evaluate the masses and leptonic widths of the bottomonium and charmonium meson states using the Hamiltonian in Eq. (\ref{eqn:ham}) and the leptonic widths formula in Eq. (\ref{eqn:DW}). The theoretical results for the $1^{--}$ $1S$ to $5D$ states are listed in Table~\ref{tab:1--md}, with $M^{cal}$ for masses and $\Gamma^{cal}$ for leptonic widths. The experimental data $M^{exp}$ and $\Gamma^{exp}$ of $\Upsilon$(1S) to $\Upsilon$(4S) and $\psi$(1S), $\psi$(2S), $\psi$(4040), and $\psi$(3770) are taken from PDG~\cite{PDG}. These states are widely believed to be conventional meson states.

For comparison, we also briefly discuss the results of several works reviewed in Sec.~\ref{sec:Int}, and show their predictions in Table~\ref{tab:1--md}. 

For bottomonium states, the fitting results of masses~\cite{Shah:2012js} can be matched very well with experimental data, but the leptonic widths are all smaller than experimental data especially for $\Upsilon$(2S), $\Upsilon$(3S), and $\psi$(2S).

The theoretical mass results of $1^{--}$ bottomonium states from Refs~\cite{Godfrey:2015dia,Wang:2018rjg} are roughly compatible with experimental data, and the mass of 3S states has a very nice match with $\Upsilon$(3S). However, both of leptonic width results are significantly larger than experimental data from $\Upsilon$(1S) to $\Upsilon$(4S).

On the other hand, mass results in Ref~\cite{Segovia:2016xqb} are roughly compatible but leptonic width results are significantly smaller than the data.

For charmonium states, the collected theoretical results of Ref~\cite{Li:2009zu} show that the 1S mass agrees well with the data of $J/\psi$, and the masses of 2S, 1D, and 3S are compatible with the data of $\psi$(2S), $\psi$(3770), and $\psi$(4040). But theoretical leptonic width results are all larger than the corresponded data. The results of Ref~\cite{Segovia:2008zz} show that the theoretical masses are roughly compatible with the data, but the leptonic widths of 1S and 2S states are much smaller than the data.

The theoretical mass results~\cite{Barnes:2005pb} are compatible with the data, but the leptonic width results~\cite{Barnes:2004cz} are too large due to only the leading order contribution in leptonic width formula considered.

It can be seen from Table~\ref{tab:1--md} that the predictions of mass and leptonic width for higher excited $1^{--}$ states do not simultaneously match well with experimental data when one considers the meson states in either S-wave or D-wave state only.

\subsection{\label{sec:mix} Possible mixtures of $nS$ and $(n-1)D$ states}

\begin{table*}[tb]
\caption{The mixtures of $nS$ and $(n-1)D$ $1^{--}$ charmonium and bottomoium states. $M^{exp}$ and $\Gamma^{exp}$ are from PDG~\cite{PDG}.}
\label{tab:1--mix}
\begin{ruledtabular}
\begin{tabular}{ccccccccc}
quark & mixture &  \multirow{2}{*}{$M^{cal}  {\rm (MeV)}$} & \multirow{2}{*}{$\theta\degree$} & $M_{\psi_1}$  &  \multirow{2}{*}{Assignment} & \multirow{2}{*}{$M^{exp}  {\rm (MeV)}$} & $\Gamma_{\psi_1}$ & \multirow{2}{*}{$\Gamma^{exp}  {\rm (keV)}$}
\\
content & states & & & $M_{\psi_2}$  & & & $\Gamma_{\psi_2}$ & 
\\
\hline

\multirow{2}{*}{$b\bar b$}& 2S & 10017 & \multirow{2}{*}{$-9.0\degree$} \multirow{2}{*}{,} \multirow{2}{*}{$15.1\degree$} & 10014\,, 10007 & $\Upsilon(2S)$ & $10023.26\pm0.31$  & 0.601 & $0.612\pm0.011$
\\
& 1D & 10143 & & 10146\,, 10153 & ... & ... & 0.027 & ...
\\

\multirow{2}{*}{$b\bar b$}& 3S & 10379 & \multirow{2}{*}{$-12.5\degree$} \multirow{2}{*}{,} \multirow{2}{*}{$22.2\degree$} & 10375\,, 10363 & $\Upsilon(3S)$ & $10355.2\pm0.5$  & 0.430 & $0.443\pm0.008$
\\
& 2D & 10461 & & 10465\,, 10477 & ... & ... & 0.042 & ...
\\
\multirow{2}{*}{$b\bar b$}&4S&10678 & \multirow{2}{*}{$38.0\degree$}\multirow{2}{*}{,} \multirow{2}{*}{$-25.3\degree$}  & 10583\,, 10661 & $\Upsilon(4S)$ & $10579.4\pm1.2$  & 0.288 & $0.272\pm0.029$ 
\\
&3D&10739 & & 10834\,, 10756 & $\Upsilon(10753)$? & $10753\pm6$ & 0.109 & ...
\\
\multirow{2}{*}{$b\bar b$}&5S&10942 & \multirow{2}{*}{$34.9\degree$}\multirow{2}{*}{,}  \multirow{2}{*}{$-19.6\degree$}  & 10897\,, 10935 & $\Upsilon(10860)$ & $10885.2^{+2.6}_{-1.6}$ & {0.278} & $0.31\pm0.07$ 
\\
&4D& 10991 & & 11036\,, 10998 & $\Upsilon(11020)$ & $11000\pm4$ & 0.074 & $0.13\pm0.03$ 
\\
\hline
\multirow{2}{*}{$c\bar c$}&2S& 3673 & \multirow{2}{*}{$-2.5\degree$} \multirow{2}{*}{,} \multirow{2}{*}{$30.6\degree$} & 3673\,, 3615 & $\psi(2S)$ & $3686.10\pm0.03$ & 2.27 & $2.33\pm0.04$  
\\
&1D& 3782 & & 3782\,, 3840 & $\psi(3770)$ & $3773.13\pm0.35$ & 0.20 & $0.26\pm0.02$ 
\\
\multirow{2}{*}{$c\bar c$}&3S&4046 & \multirow{2}{*}{$-21.2\degree$} \multirow{2}{*}{,} \multirow{2}{*}{$62.6\degree$} & 4034\,, 4139 & $\psi(4040)$ & $4039\pm1$ & 0.98 & $0.86\pm0.07$  
\\
&2D&4114 & & 4125\,, 4021 & $\psi(4160)$ & $4191\pm5$ & 0.79 & $0.48\pm0.22$ 
\\
\multirow{2}{*}{$c\bar c$}&4S & 4355 & \multirow{2}{*}{$-18.1\degree$} \multirow{2}{*}{,} \multirow{2}{*}{$68.3\degree$} & 4349\,, 4413  & $\psi(4360)$ & $4368\pm13$ & 0.77 & ...  
\\
&3D& 4404 & & 4410\,, 4346  & $\psi(4415)$ & $4421\pm4$ & 0.68 & $0.58\pm0.07$
\\
\end{tabular}
\end{ruledtabular}
\end{table*}
As reviewed above, it is difficult to simultaneously reproduce masses and leptonic widths of experimental data for higher excited quarkonia under the assumption of pure S- and D-wave states. 

Based on the results in Table~\ref{tab:1--md}, it is natural to consider altering the theoretical masses and leptonic widths simultaneously by mixing the S and D waves. Dynamically, the coupling of S and D waves may stem from tensor forces.  However, detailed calculations reveal that the tensor force in the Hamiltonian in Eq. (\ref{eqn:ham}) and (\ref{eqn:BF}) is not strong enough to mix the $S$ and $D$ waves considerably.
{One may expect that coupled-channel effects, stemming from the mixing via decay channels, might be the source of the large S-D mixing~\cite{Heikkila:1983wd, Ono:1985eu, Ono:1985jt, Lu:2016mbb, Fu:2018yxq}. Or, the meson exchange or multi-gluon exchange may contribute stronger tensor force interactions~\cite{Machleidt:1987hj, Glozman:1995fu}. }

The mixing probability is proportional to $1/|\delta E|^2$ in perturbation calculations, where $\delta E$ is the energy difference between the two mixed states, and hence only the nearest states may mix up considerably. Based on the results in Table~\ref{tab:1--md}, we estimate that the probability for the $(n-2)D$ and $nS$ mixing as well as the $nD$ and $nS$ mixing is less than 10\% of the probability for the $nS$ and $(n-1)D$ mixing. It is a reasonable approximation to consider only the mixing between the nearest $nS$ and $(n-1)D$ states.
The mixed states may take the form,
\begin{flalign}\label{eqn:mwf}
&|\psi_1\rangle={\rm{cos}}\theta|nS\rangle+{\rm{sin}}\theta|(n-1)D\rangle, \nonumber \\
&|\psi_2\rangle=-{\rm{sin}}\theta|nS\rangle+{\rm{cos}}\theta|(n-1)D\rangle, 
\end{flalign}
where $\theta$ is mixing angle.

The charmonium states, $\psi(2S)$, $\psi(3770)$, $\psi(4040)$, $\psi(4160)$, $\psi(4360)$ and $\psi(4415)$, and bottomonium states $\Upsilon(2S)$, $\Upsilon(3S)$, $\Upsilon(4S)$, $\Upsilon(10860)$, and $\Upsilon(11020)$ are considered to be $S$-$D$ mixture candidates. The masses, $M_{\psi_1}$ and $M_{\psi_2}$, and the leptonic decay widths, $\Gamma_{\psi_1}$ and $\Gamma_{\psi_2}$, of the states $|\psi_1\rangle$ and $|\psi_2\rangle$ are derived,
\begin{flalign}\label{eqn:MG}
&M_{\psi_1}=\frac12(M_{nS}+M_{(n-1)D}+(M_{nS}-M_{(n-1)D})\frac1{{\rm{cos}}2\theta}), \nonumber \\
&M_{\psi_2}=\frac12(M_{nS}+M_{(n-1)D}+(M_{(n-1)D}-M_{nS})\frac1{{\rm{cos}}2\theta}), \nonumber \\
&\Gamma_{\psi_1}=(\sqrt{F_c}F_{nS}R_{nS}(0){\rm{cos}}\theta+F_{(n-1)D}R^{''}_{(n-1)D}(0){\rm{sin}}\theta)^2, \nonumber \\
&\Gamma_{\psi_2}=(-\sqrt{F_c}{F_{nS}R_{nS}(0)\rm{sin}}\theta+F_{(n-1)D}R^{''}_{(n-1)D}(0){\rm{cos}}\theta)^2
\end{flalign}
with
\begin{flalign}\label{eqn:F}
&F_{nS}=\frac{2\alpha e_q}{M_{nS}},\,F_{nD}=\frac{5\alpha e_q}{\sqrt 2m_i^2M_{nD}},\,F_{c}=(1-\frac{16\alpha_s}{3\pi}).
\end{flalign}

Fitting the theoretical leptonic widths ($\Gamma_{\psi_1}$ and $\Gamma_{\psi_2}$) of the $S$-$D$ mixture states to experimental data leads to two mixing angles $\theta\degree$, as shown in the 4th column of Table~\ref{tab:1--mix}. 
By applying the two angles to Eq.~(\ref{eqn:MG}), we derive two masses for each mixing state shown in the 5th column.
{It is found that the masses derived with the 1st angle in column 4 are more consistent with experimental data. }

{The decay widths of E1 radiative transitions are calculated for the $S$-$D$ mixture states, since the radiative transitions are sensitive to the internal structure of states. The decay width for E1 transitions between an initial state $n^{2S+1}L_J$ and final state $n'^{2S'+1}L'_{J'}$ can be written as\cite{Segovia:2016xqb}}
\begin{flalign}\label{eqn:E1}
&\Gamma(n^{2S+1}L_J\to n'^{2S'+1}L'_{J'}+\gamma)  \nonumber\\
&=\frac{4\alpha^2e_q^2k^3}{3}(2J'+1)S^{E}_{fi}\delta_{SS'}|\epsilon_{fi}|^2\frac{E_f}{M_i},  
\end{flalign}

\begin{flalign}
S^{E}_{fi}={\rm max}(L,L'){\begin{Bmatrix}
J&1&J' \\
L'&S&L
\end{Bmatrix}}^2,  
\end{flalign}

\begin{flalign}\label{eqn:WF}
\epsilon_{fi}=\frac3k\int_0^\infty R_i(r)\left[\frac{kr}{2}j_0\left(\frac{kr}2\right)-j_1\left(\frac{kr}2\right)\right]R_f(r)r^2dr
\end{flalign}
{where k is the emitted photon momentum. $M_i$ is the mass of the initial state and $E_f$ is the energy of the final state, which are taken from established experimental data. 
$S^{E}_{fi}$ is a the statistical factor. $j_i(x)$ is the spherical Bessel functions of the first kind. $R_i(r)$ and $R_f(r)$ are the radial wave function of initial and final states respectively.}
 
{Since the E1 transition branching fractions in PDG of $\Upsilon$(2S), $\Upsilon$(3S), $\psi$(2S) and $\psi$(3770) are clear, 
E1 transition decay processes, $\Upsilon \rm{(2S-1D)}\to\gamma \chi_{b_J}\rm{(1P)}$, $\Upsilon \rm{(3S-2D)}\to\gamma \chi_{b_J}\rm{(2P)}$ and $\psi \rm{(2S-1D)}\to\gamma \chi_{b_J}\rm{(1P)}$, are studied.
The E1 decay widths are calculated by applying the $S$-$D$ mixed radial wave function Eq.~(\ref{eqn:mwf}) for initial states to Eq.~(\ref{eqn:WF}). $M_i$ in Eq.~(\ref{eqn:E1}) for $\Upsilon$ D-wave mixture states are taken from the mass spectrum in Table~\ref{tab:1--mix} due to no available data.
The theoretical results, compared with experimental data, are listed in Table~\ref{tab:E1}. In this case, experimental data of E1 decay widths is derived from the experimental data of total decay widths and E1 branching fractions of PDG 2022~\cite{PDG}.
}

\begin{table}[tb]
\caption{Theoretical results and experimental data of E1 radiative transition decay widths of $\Upsilon$ and $\psi$ mixture states.}
\label{tab:E1}
\begin{ruledtabular}
\begin{tabular}{ccccccccc}
\splitcell{Initial\\state} & \splitcell{Final\\state} & \splitcell{$\mathcal{B} ^{exp}_{E1}$~\cite{PDG}\\($\Gamma_{E1}/\Gamma _{tot}$)} & \splitcell{$\Gamma^{exp} _{E1}$\\$(\rm{keV})$} & \splitcell{$\Gamma^{the} _{E1}$\\$(\rm{keV})$}
\\
\hline
\bsplitcell{$\Upsilon$(2S)\\$\Upsilon$(1D)} & $\gamma\chi_{b0}(1P)$ &\splitcell{$(3.8\pm0.4)\%$\\...}& \splitcell{$1.2\pm0.2$\\...} & \splitcell{1.0\\8.2}                               
\\
\bsplitcell{$\Upsilon$(2S)\\$\Upsilon$(1D)}   & $\gamma\chi_{b1}(1P)$ &\splitcell{$(6.9\pm0.4)\%$\\...}& \splitcell{$2.2\pm0.3$\\...} & \splitcell{1.8\\4.8}                               
\\
\bsplitcell{$\Upsilon$(2S)\\$\Upsilon$(1D)}   & $\gamma\chi_{b2}(1P)$ &\splitcell{$(7.15\pm0.35$)\%\\...}& \splitcell{$2.3\pm0.3$\\...} & \splitcell{2.1\\0.3}                               
\\
\\
\bsplitcell{$\Upsilon$(3S)\\$\Upsilon$(2D)} & $\gamma\chi_{b0}(2P)$ &\splitcell{$(5.9\pm0.6)\%$\\...}& \splitcell{$1.2\pm0.2$\\...} & \splitcell{1.1\\5.7}                                   
\\
\bsplitcell{$\Upsilon$(3S)\\$\Upsilon$(2D)}  & $\gamma\chi_{b1}(2P)$ &\splitcell{$(12.6\pm1.2)\%$\\...}& \splitcell{$2.6\pm0.5$\\...} & \splitcell{2.1\\3.8}                                   
\\
\bsplitcell{$\Upsilon$(3S)\\$\Upsilon$(2D)}  & $\gamma\chi_{b2}(2P)$ &\splitcell{$(13.1\pm1.6)\%$\\...}& \splitcell{$2.7\pm0.6$\\...} & \splitcell{2.4\\0.2}                                   
\\
\\
\bsplitcell{$\psi$(2S)\\$\psi$(1D)} & $\gamma\chi_{c0}(1P)$ &\splitcell{$(9.79\pm0.20)\%$\\$(0.69\pm0.06)\%$}& \splitcell{$28.8\pm1.4$\\$187.7\pm23.8$} & \splitcell{24.6\\138.9}
\\
\bsplitcell{$\psi$(2S)\\$\psi$(1D)}&  $\gamma\chi_{c1}(1P)$ &\splitcell{$(9.75\pm0.24)\%$\\$(0.249\pm0.023)\%$}& \splitcell{$28.7\pm1.5$\\$67.7\pm9.0$} & \splitcell{35.7\\65.3} 
\\
\bsplitcell{$\psi$(2S)\\$\psi$(1D)}&  $\gamma\chi_{c2}(1P)$ &\splitcell{$(9.52\pm0.20)\%$\\$<6.4\times10^{-4}$}& \splitcell{$28.0\pm1.4$\\$<17.4$} & \splitcell{32.7\\3.1}                                    
\end{tabular}
\end{ruledtabular}
\end{table}

\subsection{\label{sec:AD} Assignments and discussion}

The theoretical mass and leptonic width results of $1^{--}$ heavy quarkonium states are summarized in Table~\ref{tab:1--assi}, where some possible $S$-$D$ mixing states are listed in brackets, and the tentative assignments for the observed states are provided. 

\begin{table*}[htbp]
\caption{Present predictions of bottomoium and charmonium $1^{--}$ state masses (MeV) and leptonic widths (keV) after possible $S$-$D$ mixture compared with experimental data. The experimental data is taken from PDG~\cite{PDG}.}
\label{tab:1--assi}
\begin{ruledtabular}
\begin{tabular}{ccccccccc}
 $nL$ & $M^{cal}_{S-D}{\rm (MeV)}$ & Assignment & $M^{exp}{\rm (MeV)}$ & $\Gamma^{cal}_{S-D}{\rm (keV)}$ & $\Gamma^{exp}{\rm (keV)}$  &  Other assignments
 \\
\hline
1S     & 9461   & $\Upsilon(1S)$       & $9460.30\pm0.26$  & 1.370 & $1.340\pm0.018$ & 1S $b\bar b$~\cite{PDG}
\\
\bsplitcell{2S\\1D} & \bsplitcell{10014\\10146} & \splitcell{$\Upsilon(2S)$\\...}  &  \splitcell{$10023.26\pm0.31$\\...}   & \splitcell{0.601\\0.027}  & \splitcell{$0.612\pm0.011$\\...} & \splitcell{2S $b\bar b$~\cite{PDG}\\...}
\\
\bsplitcell{3S\\2D} & \bsplitcell{10375\\10465} & \splitcell{$\Upsilon(3S)$\\...}  &  \splitcell{$10355.2\pm0.5$\\...}   & \splitcell{0.430\\0.042}  & \splitcell{$0.443\pm0.008$\\...} & \splitcell{3S $b\bar b$~\cite{PDG}\\...}
\\
\bsplitcell{4S\\3D} & \bsplitcell{10583\\10834} & \splitcell{$\Upsilon(4S)$\\$\Upsilon(10753)$?}  &  \splitcell{$10579.4\pm1.2$\\$10753\pm6$}   & \splitcell{0.288\\0.109}  & \splitcell{$0.272\pm0.029$\\...} & \splitcell{4S $b\bar b$~\cite{PDG}\\...}
\\
\bsplitcell{5S\\4D} & \bsplitcell{10897\\11036} & \splitcell{$\Upsilon(10860)$\\$\Upsilon(11020)$} & \splitcell{$10885.2^{+2.6}_{-1.6}$\\$11000\pm4$} & \splitcell{0.278\\0.074} & \splitcell{$0.31\pm0.07$\\$0.13\pm0.03$}  
& \splitcell{5S $b\bar b$~\cite{Segovia:2016xqb,Wang:2018rjg,Godfrey:2015dia,Deng:2016ktl,Dong:2020tdw}\\6S $b\bar b$~\cite{Segovia:2016xqb,Wang:2018rjg,Godfrey:2015dia,Deng:2016ktl,Dong:2020tdw},7S $b\bar b$~\cite{Shah:2012js}}
\\
&&&&&& 
\\
1S     & 3110   & $J/\psi$                   & $3096.90\pm0.01$  & 6.02    &  $5.55\pm0.14$  & 1S $c\bar c$~\cite{PDG}
\\
\bsplitcell{2S\\1D}  & \bsplitcell{3673\\3782}  &  \splitcell{$\psi(2S)$\\$\psi(3770)$}  & \splitcell{$3686.10\pm0.03$\\$3773.13\pm0.35$}  & \splitcell{2.27\\0.20}  & \splitcell{$2.33\pm0.04$\\$0.26\pm0.02$}   & \splitcell{2S $c\bar c$~\cite{PDG}\\1D $c\bar c$~\cite{HadronSpectrum:2012gic,Zhang:2006td,Li:2009zu,Segovia:2008zz,Barnes:2005pb}}
\\
\bsplitcell{3S\\2D}  & \bsplitcell{4034\\4125}  &  \splitcell{$\psi(4040)$\\$\psi(4160)$}  & \splitcell{$4039\pm1$\\$4191\pm5$}  & \splitcell{0.98\\0.79}  & \splitcell{$0.86\pm0.07$\\$0.48\pm0.22$}   & \splitcell{3S $c\bar c$~\cite{Zhang:2006td,Li:2009zu,Segovia:2008zz,Barnes:2005pb}\\2D $c\bar c$~\cite{Zhang:2006td,Li:2009zu,Segovia:2008zz,Barnes:2005pb}}
\\
 ...     & ...       &  $\psi(4230)$           & $4230\pm8$            &   ...     &   ...                        & 4S $c\bar c$~\cite{Llanes-Estrada:2005qvr, Li:2009zu, Shah:2012js}, 3D $c\bar c$~\cite{Zhang:2006td,Dai:2012pb},
\\     
         &           &                                 &                                 &            &                             & $c\bar cg$~\cite{HadronSpectrum:2012gic,Luo:2005zg,Zhu:2005hp}, $\left(qc\bar q\bar c\right)$~\cite{Maiani:2005pe,Drenska:2009cd,Ebert:2008kb}, $\left(q\bar c\right)\left(\bar qc\right)$~\cite{Chiu:2005ey,Ding:2008gr,Close:2009ag,Close:2010wq}
\\
\bsplitcell{4S\\3D}  &  \bsplitcell{4349\\4410}  &   \splitcell{$\psi(4360)$\\$\psi(4415)$}  &   \splitcell{$4368\pm13$\\$4421\pm4$}  & \splitcell{0.77\\0.68}  &   \splitcell{...\\$0.58\pm0.07$}  &  \splitcell{4S $c\bar c$~\cite{Segovia:2008zz}, 3D $c\bar c$~\cite{Li:2009zu}\\4S $c\bar c$~\cite{Zhang:2006td}, 3D $c\bar c$~\cite{Segovia:2008zz}, 5S $c\bar c$~\cite{Li:2009zu,Shah:2012js}}
\\
5S    & 4628   & $\psi(4660)$            & $4643\pm9$           &     0.97     &         ...            & 5S $c\bar c$~\cite{Segovia:2008zz,Ding:2007rg}, 6S $c\bar c$~\cite{Li:2009zu,Shah:2012js}
\\
4D    & 4667  & ...  & ...    &  0.20 & ...
\\
\end{tabular}
\end{ruledtabular}
\end{table*}

For excited bottomonium states 2S--1D, 3S--2D, 4S--3D, and 5S--4D mixtures are considered. The $\Upsilon$(10023) and $\Upsilon$(10355) are assigned to be largely 2S and 3S state respectively, containing some D wave component. The $\Upsilon$(10579) is assigned a 4S--3D mixture state due to the large mixing angle.

The leptonic width data of $\Upsilon$(11020), $0.13\pm0.03$ keV~\cite{PDG}, is averaged from 
$0.095\pm0.03\pm0.035$ keV~\cite{CLEO:1984vfn} and $0.156\pm0.040$ keV~\cite{Lovelock:1985nb}, which is too small to be 5S state where the 5S leptonic width is predicted to be around 0.3 keV in
Table~\ref{tab:1--md}. Thus, the $\Upsilon$(10860) and $\Upsilon$(11020) are assigned to be 5S--4D mixed states due to a congruent matching for both masses and leptonic widths. 

The newly reported state $\Upsilon$(10753) observed by Belle~\cite{Belle:2019cbt} and Belle-II collaboration~\cite{Belle-II:2022xdi} is tentatively assigned to be largely 3D state. For a tetraquark mixture interpretation, one may refer to Ref~\cite{Ali:2019okl}.
More experimental data and theoretical works are essential for making an unambiguous assignment for the $\Upsilon$(10753).

For the higher excited charmonium states, 2S--1D, 3S--2D, and 4S--3D mixtures are considered. It is found that the $\psi$(2S) possesses a small D-wave component, and $\psi$(3770) possesses a small S-wave component, in consistent with our previous work~\cite{Sreethawong:2014jra, Limphirat:2013jga} and other theoretical work~\cite{Li:2009zu}.

Since the theoretical results of $\psi(4040)$ leptonic width (from 0.96--3.48 keV) in Table~\ref{tab:1--md} are all larger than experimental data~\cite{PDG} (with $\Gamma_{ee}=0.86\pm0.07$ keV) significantly, and the leptonic width of the widely believed 2D state $\psi(4160)$~\cite{Zhang:2006td,Li:2009zu,Segovia:2008zz,Barnes:2005pb} is measured to be $0.48\pm0.22$ keV~\cite{BES:2007zwq}, one may naturally consider the $\psi(4040)$ and $\psi(4160)$ to be S-D mixture states. The PDG mass, $4191\pm5$ MeV, of the $\psi(4160)$~\cite{PDG}is collected from BES collaboration~\cite{BES:2007zwq}. However, data analyses in Ref.~\cite{Seth:2004py} result in the mass and leptonic width, $4151\pm4$ MeV and $0.83\pm0.08$ keV from Crystal Ball measurement~\cite{Osterheld:1986hw}, and $4155\pm5$ MeV and $0.84\pm0.13$ keV from BES measurement~\cite{BES:2001ckj}. Our theoretical results are compatible with the results in Ref.~\cite{Seth:2004py}, and we suggest that the $\psi(4040)$ and $\psi(4160)$ are 3S and 2D mixed states.  

In other conventional meson assignments, the $\psi(4360)$ is assigned to be 4S $c\bar c$~\cite{Segovia:2008zz} and 3D $c\bar c$~\cite{Li:2009zu} while the $\psi$(4415) is assigned to be 4S $c\bar c$~\cite{Zhang:2006td}, 3D $c\bar c$~\cite{Segovia:2008zz} and 5S $c\bar c$~\cite{Li:2009zu,Shah:2012js}. Considering the congruent matching for both masses and leptonic widths in the work, we assign the $\psi(4360)$ and $\psi(4415)$ to be 4S and 3D mixture states, where the $\psi(4360)$ and $\psi(4415)$ are largely 4S and 3D state respectively. 

$\psi(4660)$ is tentatively assigned to be 5S state according to the good mass matching, which is consistent with the Refs~\cite{Segovia:2008zz,Ding:2007rg}. The $\psi(4230)$ can not be accommodated as a $c\bar c$ state in the present work. For other interpretations, one may refer to Refs~\cite{HadronSpectrum:2012gic,Luo:2005zg,Zhu:2005hp} for charmonium hybrid, Refs~\cite{Maiani:2005pe,Drenska:2009cd,Ebert:2008kb} for tetraquark, and Refs~\cite{Chiu:2005ey,Ding:2008gr,Close:2009ag,Close:2010wq} for molecule picture.

\section{Summary}\label{sec:SUM}

The masses and leptonic decay widths of S-wave and D-wave heavy quarkonium meson states with quantum number $J^{PC}=1^{--}$ until 6S and 5D have been evaluated, with all model parameters predetermined by studying all ground and first radial excited S- and P-wave heavy quarkonium mesons. The theoretical results have been matched with experimental data by considering possible S-D mixtures, and the tentative assignments for higher excited states are provided. {Based on the assignment, E1 radiative transition decay widths are calculated.}

For the $1^{--}$ bottomonium states, this work suggests that the $\Upsilon(2S)$ and $\Upsilon(3S)$ may possesses some D-wave component, and $\Upsilon(4S)$ may be a 4S-3D mixture state. The $\Upsilon$(10860) and $\Upsilon$(11020) are assigned to be 5S-4D mixture states. The $\Upsilon$(10753) is tentatively assigned to be 4S-3D mixture state, and more experimental data is required to make unambiguous assignment for this newly reported state.

For the $1^{--}$ charmonium states, the work suggests that the $\psi(2S)$ and $\psi(3770)$ may possesses some small D-wave and S-wave component respectively, and the $\psi$(4040) and $\psi$(4160) are mainly 3S and 2D state respectively. The $\psi$(4360) and $\psi$(4415) are largely 4S and 3D state respectively. The $\psi$(4660) is assigned to be a 5S state. The $\psi$(4230) may not be accommodated with the conventional meson picture in the present work.

{The work shows that a large $S-D$ mixing is essential to understand the experimental data of higher excited quarkonia, but the tensor force in the widely applied Hamiltonian is not strong enough to mix the $S$ and $D$ waves considerably. It is expected that the coupled-channel effects, resulting from couplings to common decay channels, might be an important source of the large S-D mixing.  Heavy quarkonia will be studied by considering the coupled channel induced S-D mixing in our future work.}

\begin{acknowledgments}
This work was supported by (i) Suranaree University of Technology (SUT), (ii) Thailand Science Research and Innovation (TSRI), and (iii) National Science Research and Innovation Fund (NSRF), Project No. 179349.
\end{acknowledgments}

\bibliography{PRD2022}

\end{document}